\documentclass[useAMS,usenatbib]{mn2e}

\include{epsf}

\def\gsim{\lower.73ex\hbox{$\sim$}\llap{\raise.4ex\hbox{$>$}}$\,$}
\def\lsim{\lower.73ex\hbox{$\sim$}\llap{\raise.4ex\hbox{$<$}}$\,$}
\def\etal{{\it et al}.}
\def\kms{km s$^{-1}$}
\def\Mpch{\, h^{-1}{\rm Mpc}}    %%% ~ creates a punctuation problem
\def\kms{\, {{\rm km}}\,{{\rm s}}^{-1} }
\def\be{\begin{equation}}
\def\ee{\end{equation}}

\title[$z$-Space Distortions in  LBG surveys]
  {Constraining $\beta(z)$ and $\Omega_{m}^{0}$ from Redshift-Space Distortions in $z\sim 3$ galaxy surveys}
\author[J. da \^{A}ngela, P. J. Outram, T. Shanks]
  {J. da \^{A}ngela$^{1,2}$, 
  P. J. Outram$^1$, T. Shanks$^1$ \\
  1 Department of Physics, University of Durham,Science Laboratories,South Road, Durham, DH1 3LE, United Kingdom\\
  2 Centro de Astrof\'{\i}sica da Universidade do Porto, R. das Estrelas s/n, 4150-762 Porto, Portugal}

\begin{document}

\pagerange{\pageref{firstpage}--\pageref{lastpage}} \pubyear{2005}

\maketitle

\label{firstpage}

\begin{abstract}

We use sample of 813 Lyman-break galaxies (LBGs) with $2.6<z<3.4$ to perform a detailed analysis of the redshift-space ($z$-space) distortions in their clustering pattern and from them derive confidence levels in the $[\Omega_{m}^{0},\beta(z=3)]$ plane. We model the $z$-space distortions in the shape of the correlation function measured in orthogonal directions, $\xi(\sigma,\pi)$. This modeling requires an accurate description of the real-space correlation function to be given as an input. From the projection of $\xi(\sigma,\pi)$ in the angular direction, $w_{p}(\sigma)$, we derive the best fitting amplitude and slope for the LBG real-space correlation function:  $r_{0}=4.48^{+0.17}_{-0.18} \Mpch$ and $\gamma=1.76^{+0.08}_{-0.09}$ ($\xi(r)= (r/r_{0})^{-\gamma}$). A comparison between the shape of $\xi(s)$ and $w_{p}(\sigma)$ suggests that $\xi(r)$ deviates from a simple power-law model, with a break at $\sim 9 \Mpch$. This model is consistent with the observed projected correlation function. However, due to the limited size of the fields used, the $w_{p}(\sigma)$ results are limited to $\sigma \lsim 10 \Mpch$. Assuming this double power-law model, and by analysing the shape distortions in $\xi(\sigma,\pi)$, we find the following constraints: $\beta(z=3) = 0.15^{+0.20}_{-0.15}$, $\Omega_{m}^{0} = 0.35^{+0.65}_{-0.22}$. Combining these results with orthogonal constraints from linear evolution of density perturbations, we find that $\beta(z=3) = 0.25^{+0.05}_{-0.06}$, $\Omega_{m}^{0} = 0.55^{+0.45}_{-0.16}$.

\end{abstract}


\section{Introduction}

To compute the distances to high-redshift objects, one can assume that their observed redshifts are only due to the expansion of the Universe. Then, for a given assumed cosmology, a relation between comoving distance and redshift is easily found. However, the observed redshifts do not exclusively depend on the expansion rate of the Universe, as they also include contributions from the motions of the galaxies in their local rest-frame, superimposed on the systemic flow due to the Hubble expansion. If these contributions are not taken into account, the estimated distances are said to be measured in {\it redshift-space} ($z$-space), and they do not reflect the true, comoving distances to the galaxies. If the distances could be estimated considering the effects of peculiar motions, then these would be said to be measured in {\it real-space}. 

Peculiar velocities introduce distortions in the measured clustering pattern. If a given set of galaxies has a spherically symmetric clustering pattern in real-space, but the galaxies have a very large velocity dispersion, then the clustering signal measured in $z$-space will be smeared along the line-of-sight. These features are commonly referred to as ``fingers of God''. Also, gravitational instabilities at large scales cause a coherent infall of the galaxies into the potential well of higher density regions. 

Besides the $z$-space distortions, geometric distortions in the clustering pattern also occur if a {\it wrong} cosmology is assumed to convert the measured redshifts into comoving distances. This is because the clustering measured along the line-of-sight has a dependence on cosmology that differs from that of the clustering measurements in the angular direction.

The goal of this work is to use dynamical and geometric distortions to study the large scale coherent infall and put constraints on cosmology, from a sample of $\sim 800$ Lyman-break galaxies. The infall can be quantified by measuring $\beta$, defined as $\beta = \Omega_{m}^{0.6}/b$ ($b$ being the linear bias parameter). The relation between the clustering measured in real- and $z$-space, in one dimension, in the linear regime, was derived by \citet{kaiser}. He found that the large-scale infall will simply produce an amplitude shift of the correlation function, measured in real- and $z$-space. The use of geometric distortions to study cosmology was pioneered by \citet{ap}, who  demonstrated that they can be used as a powerful cosmological test for a non-zero $\Lambda$.

Lyman-break galaxies have $z\sim 3$ and, due to their high sky density ($\sim 1$ LBG per arcmin$^2$) and scientific interest, such as the study of galaxy formation/evolution at early times in the history of the Universe or the existence/role of feedback mechanisms at $z\sim 3$, they are an excellent class of objects to study $z$-space distortions. They can be selected by identifying the Lyman continuum discontinuity which, at the galaxy's redshift, will determine their location in the $[U-B,B-R]$ plane.

The structure of this {\it paper} is as follows: in Section 2 we introduce the data used in our analysis. We then measure the two-dimensional $z$-space correlation function that we will use to fit the $z$-space distortions (Section 3). Integrating this along the redshift direction allow us to study the real-space clustering, as discussed in Section 4. Since there is not  enough information in the sky direction to use this as an amplitude input in the fitting of the $z$-space distortions, we compute the spherical average of the two-dimensional $z$-space correlation function in Section 5, where we determine the amplitude and shape of the $z$-space correlation function, in 1-D. In Section 6 we determine the constraints that can be obtained from the fitting of the $z$-space distortions and a brief description of the method is given. The discussion and conclusions from this work are presented in Section 7.

\section{The Data}

The LBGs used in this work are included in the sample of \citet{std}. The survey presented in their work comprises 17 fields, covering an area of 0.38 deg$^2$, and the total number of $2.67 < z < 3.25$ LBGs spectroscopically confirmed is 940.

The details of the complete survey and the data used can be found in \citet{std}. $[U,G,\mathcal{R}]_{AB}$ imaging was obtained at several telescopes and used to select the LBGs via the Lyman break technique. The spectroscopic follow-up was performed at both the Keck I and Keck II telescopes. The size of the largest field is $15.5$ arcmin and the smallest $3.7$ arcmin. These limited sizes compromise any clustering analysis that depends on the information on the sky direction. Seven of these fields contain a background QSO, whose spectrum can be used to probe the spatial distribution of Lyman $\alpha$ and metal systems.\\

The photometric catalogue from which the LBGs were selected includes 2347 photometrically selected candidates. These have an apparent $\mathcal{R}_{AB}$ magnitude limit of 25.5 and satisfy the colour criteria: $G_{AB}-\mathcal{R}_{AB} \le 1.2$, $(U_{(n,AB)}-G_{AB}) \ge (G_{AB}-\mathcal{R}_{AB})+1.0$. 

The determination of the redshift of these candidates was, in many cases, derived from the interstellar absorption lines of strong transitions, at $1200 - 1700 \AA$ rest-frame. In some cases, the identification of the LBGs was also possible by identifying the Lyman-$\alpha$ line. In order to reduce effects due to redshift errors that would influence our clustering analysis, especially at small scales, we decided to include in our sample only the LBGs with class 1 redshift, as defined by \citet{std}. The choice of considering only class 1 redshift LBGs and the redshift range $2.6<z<3.4$ leaves us with 813 selected galaxies. Table \ref{table:numb_lbg} shows the number of selected LBGs in each of the fields. The field names are the same as adopted by \citet{std}.

\begin{table}
\centering
\begin{tabular}{||l|c|c||} \hline
           Field Name &Dimension (arcmin$^2$)& Number of LBGs\\ \hline \hline
           Q0000-263& $3.69 \times 5.13$& 15  \\\hline
           CDFa& $8.80 \times 8.91$& 34 \\\hline
           CDFb& $9.05 \times 9.10$& 20 \\\hline
           Q0201+1120& $8.69 \times 8.72$& 21 \\ \hline
	   Q0256-000& $8.54 \times 8.46$& 42 \\ \hline
	   Q0302-003& $15.59 \times 15.71$& 40 \\ \hline
	   B20902+34& $6.36 \times 6.57$& 30 \\ \hline
	   Q0933+2854& $8.93 \times 9.28$& 58 \\ \hline
	   HDF-N& $8.62 \times 8.73$& 53 \\ \hline
	   Westphal& $15.02 \times 15.10$& 176 \\ \hline
	   Q1422+2309&$7.28 \times 15.51$ & 109 \\ \hline
	   3C 324& $6.65 \times 6.63$& 11 \\ \hline
	   SSA22a& $8.74 \times 8.89$& 50 \\ \hline
	   SSA22b& $8.64 \times 8.98$& 35 \\ \hline
	   DSF2237a& $9.08 \times 9.08$& 39 \\ \hline
	   DSF2237b& $8.99 \times 9.08$& 42 \\ \hline
	   Q2233+1341& $9.25 \times 9.25$& 38 \\ \hline

\end{tabular}
\caption{The dimensions of each of the LBG fields in this survey and the number of selected LBGs in each field. The field names are the same as adopted by \citet{std}. When the field contains a bright higher-$z$ QSO, the name of the field is the same as that of the QSO. Some of the fields are adjacent (CDFa and CDFb; SSA22a and SSA22b; DSF2237a and DSF2237b).}
\label{table:numb_lbg}
\end{table}

The Lyman-$\alpha$ line and the absorption lines are usually separated by a factor of a few hundred $\kms$, a feature that is often considered as evidence of powerful outflows from LBGs. It is prudent to account for this effect, in order to have a more precise estimation of the galaxies' redshifts. Following \citet{adel} we assume a simple outflow model as an explanation for this, where the interstellar absorption lines are produced ``in front'' of the outflow and hence are {\it blueshifted} relative to the galaxy; whereas the Lyman-$\alpha$ line is produced in the opposite side of the outflow, ``behind'' the galaxy in the observer's line of sight. Assuming this simple picture, the systemic redshifts of the LBGs can be determined as follows:

If no absorption features are easily identified and the redshift is determined from the Lyman-$\alpha$ line only, then the following correction is applied:

\be
z_{LBG} = z_{Ly\alpha}-\frac{v_{Ly\alpha}}{c},
\label{equation:zem}
\ee
where $z_{LBG}$ is the ``corrected'' redshift of the galaxy, $z_{Ly\alpha}$ the redshift measured from fitting a Gaussian to the Lyman-$\alpha$ line's profile, $v_{Ly\alpha}$ the mean velocity of the Lyman-$\alpha$ relative to the galaxy's nebular lines and $c$ is the speed of light. Following \citet{adel}, we take $v_{Ly\alpha} = 310 \kms$. 

Similarly, if the redshift is only estimated from the absorption lines, then the correction will be:

\be
z_{LBG} = z_{abs}-\frac{v_{abs}}{c},
\label{equation:zabs}
\ee
where $z_{abs}$ is the redshift measured from the centroids positions of well-defined absorption lines and $v_{abs}$ is the mean velocity of the interstellar absorption lines relative to the nebular lines. As we consider the absorption lines to be ``blueshifted'' relative to the galaxy, $v_{abs}$ will be negative. Following \citet{adel}, we take $v_{abs} = -150 \kms$.

In some cases, when both $z_{Ly\alpha}$ and $z_{abs}$ are measured, we apply the following correction \citep{adel}:

\be
z_{LBG} = \frac{z_{Ly\alpha}+z_{abs}}{2}-\frac{-0.114 \Delta v +230}{c},
\label{equation:zemzabs}
\ee
where $\Delta v = v_{Ly\alpha} - v_{abs}$.

\citet{adel}, from a subsample of the LBGs used in this work, found a value of $<\Delta v> = 614 \pm 314 \kms$ for the average separation between $v_{Ly\alpha}$ and $v_{abs}$. The ``velocity error'' of $314 \kms$ corresponds to an uncertainty of $\sim 5.6 \Mpch$ (comoving separation), that, in terms of $z$-space distortions, produces a similar effect to the small-scale galaxy velocity dispersion.

\begin{figure}
\begin{center}
\centerline{\epsfxsize = 9.0cm
\epsfbox{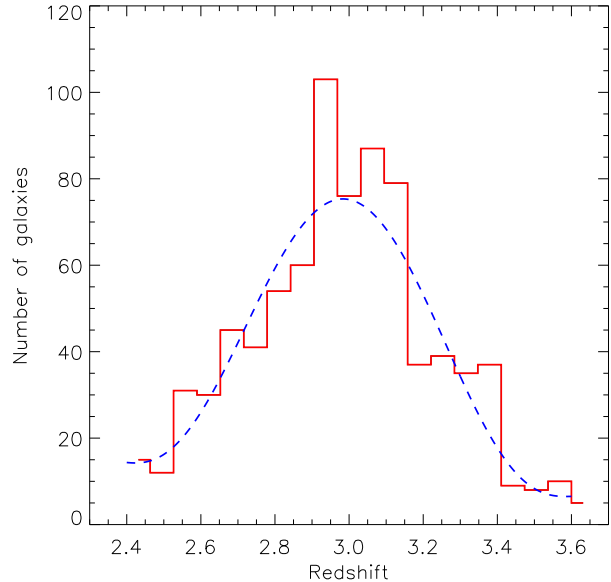}}
\caption{The solid line is an histogram showing the redshift distribution of the 813 LBGs taken from the catalogue of \citet{std}. The dashed line is a $4^{th}$ order polynomial fit to the distribution.}
\label{fig:n_z}
\end{center}
\end{figure}

Fig. \ref{fig:n_z} shows the redshift ($z_{LBG}$) distribution of our sample (solid red line), after computing the corrections discussed. This distribution is similar to the redshift distribution presented by \citet{std} (see their Fig. 10). The dashed line is a $4^{th}$ order polynomial fit to the redshift distribution.\\

To quantify the clustering, we must generate a ``random'' distribution occupying the same volume as the LBGs in the survey. In order not to wrongly interpret any completeness issue as a feature in the clustering, the random distribution must follow the same completeness as the galaxy survey. Therefore, we generated a random set of points in each the fields of the survey. We generated 20 randoms for each LBG in each field. For each of the fields, the redshift distribution of this new ensemble is described by the  $4^{th}$ order polynomial fit to the LBGs' distribution. Unfortunately, we do not have full information about the sky completeness of the individual fields. Hence, values for the declination and right ascension are randomly generated, within the field-of-view of each field. The distribution of the number of LBGs as a function of angular distance to the centre of the field (or, in some cases, the position of the background QSO) is consistent with the non-existence of radial gradients. It is possible that normalizing the random distribution to the number density of galaxies in each field causes some clustering features to be ``washed out''. If one or more specific fields lie directly in regions with particularly high (or low) clustering signal, then by generating random sets independently of the space density in other fields will cause this structure to become unnoticeable. In order to understand how significantly our measurements are affected by this we re-computed the calculations described in Sections $3$, $4$, and $5$, but using a random set of points generated for all the fields simultaneously. The difference between the two results was within the $1\ \sigma$ confidence level. Similarly, by repeating these calculations using different polynomial fits to the LBG $N(z)$ as models for the randoms' redshift distribution, we concluded that different $N(z)$ models lead to very similar clustering measurements.

\section{The redshift-space two-point correlation function, $\xi(\sigma,\pi)$}

Consider the projection of the $z$-space separation $s$, between two LBGs, along and across the line-of-sight. The $z$-space correlation function can be measured in these two directions, which are given by:
\be
\pi = |s_{2}-s_{1}|
\label{equation:pi_def}
\ee
\be
\sigma = (s_{2}+s_{1})\theta /2
\label{equation:sig_def}
\ee
where $s_{1}$ and $s_{2}$ are the distances to two different galaxies, measured in $z$-space, and $\theta$ the angular separation between them.

As mentioned above, to estimate the correlation function, a random set of points probing the same volume as the LBGs must be generated. This ensemble must have all the characteristics as the LBGs, such as the sky and redshift distributions, although it can not reproduce their clustering. 

Then, the correlation function $\xi(\sigma,\pi)$ can be computed using the estimator \citep{lsza}:

\begin{equation}
\xi(\sigma,\pi) = \frac{<DD(\sigma,\pi)>-2<DR(\sigma,\pi)>+<RR(\sigma,\pi)>}{<RR(\sigma,\pi)>},
\label{equation:xi_lasza}
\end{equation}
where $<DD(\sigma,\pi)>$ is the number of LBG-LBG pairs, $<RR(\sigma,\pi)>$ the number of random-random pairs and $<DR(\sigma,\pi)>$ the number of LBG-random pairs with separations along and across the line-of-sight given by $\sigma$ and $\pi$, respectively. 

\begin{figure}
\begin{center}
\centerline{\epsfxsize = 5.0cm
\epsfbox{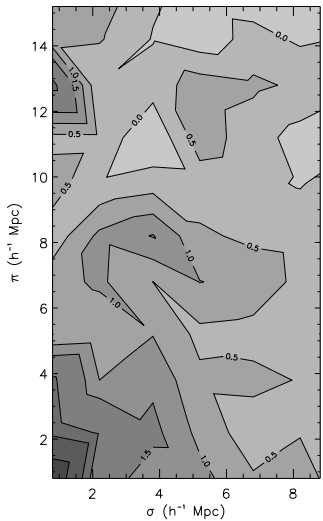}}
\caption{$\xi(\sigma,\pi)$ measured for the sample of 813 LBGs, assuming a flat, $\Omega_{m}^{0}=0.3$ cosmology.}
\label{fig:xisigpi}
\end{center}
\end{figure}

Fig. \ref{fig:xisigpi} shows $\xi(\sigma,\pi)$ measured for our LBG sample. The shape of the $\xi(\sigma, \pi)$ contours depends greatly on dynamical and geometrical effects, whose effects on the clustering can be highly anisotropic.

The LBGs' peculiar velocities lead to distortions in the $\xi(\sigma,\pi)$ shape, mainly at small scales. The random peculiar motions of the QSOs will cause an elongation of the clustering signal along the $\pi$ direction. The predominant effect at large scales is the coherent infall of the LBGs into the potential well of overdense regions, which causes a flattening of the $\xi(\sigma, \pi)$ contours along the $\pi$ direction and some elongation along $\sigma$. 

Geometric distortions also occur if the cosmology assumed to convert the observed QSO redshifts into distances is not the same as the true, underlying cosmology of the Universe. The reason is because the cosmology dependence of the separations along the redshift direction is not the same as the one of the separations measured in the perpendicular direction \citep{ap}.

Our result does not reproduce the extreme elongation along the line-of-sight seen in Fig. 20 of \citet{adel}, measured from a subsample of the LBGs used here, indicating that the feature was probably due to noise, arising from the small number of LBG pairs at small separations. The effects of the velocity error of $\sim 314 \kms$ quoted by \citet{adel} are therefore not evident in this plot, possibly due to the cancelling of "finger-of-God" effects and infall, but more probably due to the effects of noise.

\section{Obtaining the projected correlation function}

The $z$-space correlation function consists of a ``distorted'' measurement of the clustering properties of the LBGs. Our goal in this {\it paper} is to use these distortions to derive constraints on $\beta(z=3)$ and $\Omega_{m}^{0}$ and draw conclusions on the bias and infall of the galaxies at an early stage in the history of the Universe, as well as discuss the improvement that can be achieved with larger, future LBG surveys. However, the study of the real-space correlation function has, just for itself, an obvious interest, as it gives direct information about how galaxies cluster, independently of $z$-space distortion effects.

A picture of the real-space clustering can be obtained considering the clustering measured along the $\sigma$ direction, since it will not be affected by the $z$-space distortions. This can be obtained by projecting $\xi(\sigma,\pi)$ along the $\sigma$ direction, which will give information about the real-space correlation function, $\xi(r)$. Following \citet{peebles} and \citet{davis}:

\be
w_{p}(\sigma) = 2\int_{0}^{\infty} \xi(\sigma,\pi) d\pi ,
\label{equation:wp_def}
\ee
or:
\be
w_{p}(\sigma) = 2\int_{\sigma}^{\infty}\frac{r\xi(r)}{\sqrt{r^{2}-\sigma^{2}}}dr
\label{equation:wp_def_2}
\ee

\begin{figure}
\begin{center}
\centerline{\epsfxsize = 9.0cm
\epsfbox{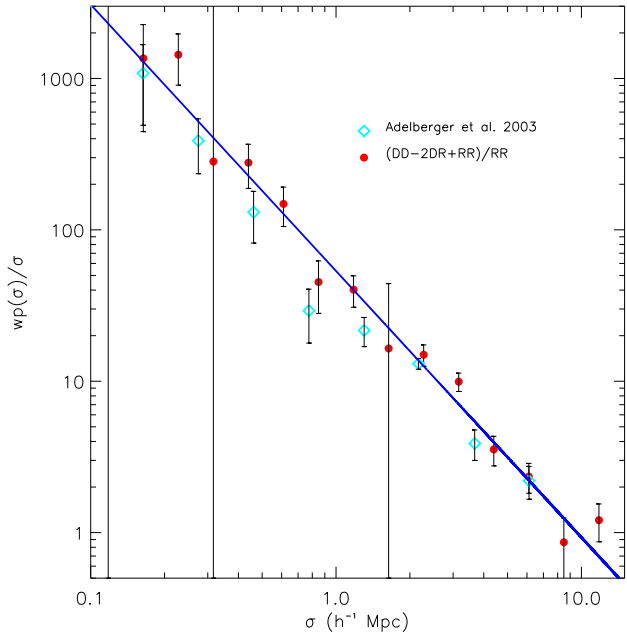}}
\caption{The projected correlation function measured for the LBG sample. The circles show the result obtained with the respective $1\sigma$ error bars. The solid line shows the best fitting power-law and the open diamonds the result obtained by \citet{adel}.}
\label{fig:wpsig}
\end{center}
\end{figure}

Fig. \ref{fig:wpsig} shows $w_{p}(\sigma)/\sigma$, obtained from integrating the already shown $\xi(\sigma,\pi)$ along the $\pi$ direction. To compute the errors, we used the Poisson estimate: $\Delta \xi = (1+\xi)\sqrt{2/<DD>}$. The circles are the measured values in the present survey. The diamonds are the values found by \citet{adel}, using a subsample of the LBG ensemble for the current work. For a better comparison with their results, we used the same values for the upper limit of the $\xi(\sigma,\pi)$ integration. Hence, this upper limit is the greater of $1000$ kms$^{-1}(1+z)/H(z)$ and $7 \sigma$. The fact that the values found by \citet{adel} are systematically below our results is probably mainly due to differences in the random catalogue generated. However, this discrepancy is smaller than the error-bars in both results.

If $\xi(r)$ can be approximated by a power-law function with the form $\xi(r)=(r/r_{0})^{-\gamma}$, it then follows from equation \ref{equation:wp_def_2} that:
\be
w_{p}(\sigma) =  r_{0}^{\gamma}\sigma^{1-\gamma}\left(\frac{\Gamma\left(\frac{1}{2}\right)\Gamma\left(\frac{\gamma-1}{2}\right)}{\Gamma\left(\frac{\gamma}{2}\right)}\right),
\label{equation:wp_sig_xi_r2}
\ee
where $\Gamma(x)$ is the Gamma function computed at $x$. Hence, $w_{p}(\sigma)/\sigma$ will also be given by a power-law with the same slope as $\xi(r)$. The best fitting power-law to the measured $w_{p}(\sigma)/\sigma$ is represented on the plot by a solid line, and it is parameterized by: $r_{0}=4.48^{+0.17}_{-0.18} \Mpch$ and $\gamma=1.76^{+0.08}_{-0.09}$. Our $\xi(r)$ measurement has a higher amplitude and it is also steeper than that found by \citet{adel} (they found $r_{0} = 3.96^{+0.29}_{-0.29}$, $\gamma=1.55^{+0.15}_{-0.15}$). \citet{fouc}, from a sample of 1294 LBG candidates from the Canada-France Deep Field Survey, measured $r_{0}$ for a fixed value of $\gamma = 1.8$ -- which is in agreement with the slope measured here. Our amplitude is still smaller than their $w(\theta)$ measurements, which indicate $r_{0}=5.9^{+0.5}_{-0.5} \Mpch$.\\

\section{Obtaining the redshift-space correlation function, $\xi(s)$}

In order to fit a model to the $z$-space distortions in $\xi(\sigma,\pi)$, the correct amplitude of the correlation function must be given as an input, in the fitting procedure. Since the fit is sensitive to the distortions in the shape of the $z$-space correlation function, its correct amplitude must be given as an input. Otherwise, the constraints obtained for $\Omega_{m}^{0}$ and $\beta(z)$ will be such that their values are those needed to compensate the input $\xi(s)$, so that the {\it amplitude} of the model $\xi(\sigma,\pi)$ fits the {\it amplitude} of $\xi(\sigma,\pi)$ from the data, rather then being a good fit to the {\it distortions} in $\xi(\sigma,\pi)$. In principle, one could use the best-fitting $\xi(r)$ power-law as an input to the $\xi(\sigma,\pi)$ model, by decomposing it in two dimensions and adding the distortions. However, due to the limited size of the fields used in the survey, the behaviour of $\xi(r)$ at scales larger than $\sim 10 \Mpch$ is unknown and, even considering that the power-law approximation is sufficiently good up to $\sim 10 \Mpch$, deviations from a simple power-law model at large scales, where $\xi(\sigma,\pi)$ is fitted, would cause shifts to the best-fitting values of $\beta(z)$ and $\Omega_{m}^{0}$. The best way of introducing the amplitude of the correlation function correctly in the $\xi(\sigma,\pi)$ model is to input a very  good description of the $z$-space correlation function's large-scale shape: $\xi(s)$. 

\begin{figure}
\begin{center}
\centerline{\epsfxsize = 9.0cm
\epsfbox{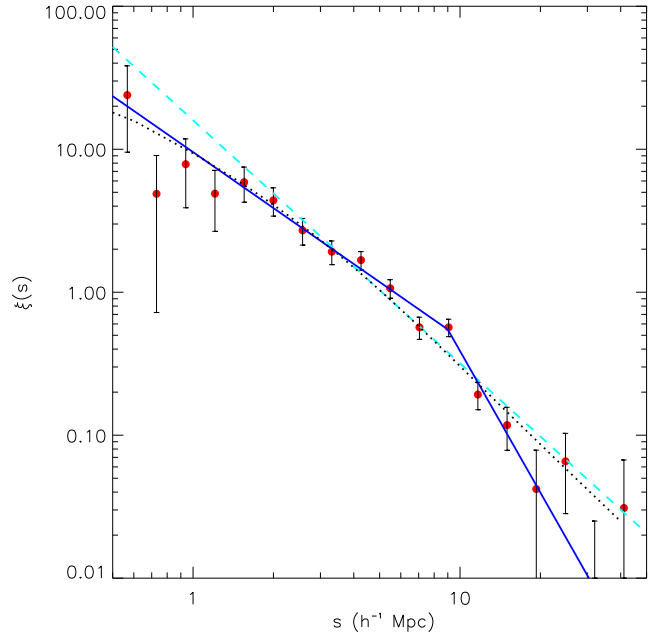}}
\caption{The $z$-space correlation function measured for the LBG sample. The circles show the result obtained with the respective $1\sigma$ Poisson error bars. The dashed line shows the best-fitting power-law to the data and the solid line the best fitting double power-law model. The dotted line is the predicted $\xi(s)$, derived from the power-law $\xi(r)$ model that best describes the $w_{p}(\sigma)$ data.}
\label{fig:xis}
\end{center}
\end{figure}

$\xi(s)$ reflects the spherical average of $\xi(\sigma,\pi)$, since $s=\sqrt{\pi^{2}+\sigma^{2}}$. Fig. \ref{fig:xis} shows our $\xi(s)$ measurements from the data.

To avoid underestimating the errors from contamination of correlated pairs of galaxies in the same separation bin, when the number of pairs is larger than the total number of galaxies the ratio $2/<DD>$ in the Poisson error estimate is replaced by $1/N_{gal}$, $N_{gal}$ being the total number of galaxies in the survey \citep{sboyle}. The circles show the measured $\xi(s)$. The error bars represent the $1 \sigma$ confidence level. The dashed line represents the best fitting power-law model to $\xi(s)$. Considering that the line is parameterized by $(s/s_{0})^{-\gamma}$, then the best fitting values of $s_{0}$ and $\gamma$ are given by $s_{0}=5.1^{+0.2}_{-0.1} \Mpch$ and $\gamma=1.71^{+0.06}_{-0.09}$. Given the observed deviations from a simple power-law model on the $\xi(s)$ shape (the reduced $\chi^{2}_{min}$ of the fit is 4.07, with 16 degrees of freedom), a double power-law model for $\xi(s)$ was also fitted to the data. It was found that the best fitting model corresponded to having the two power-laws joining at $9 \Mpch$. Then, the amplitude of the power-law probing the large scales and the slope of both the power-laws were fitted. The amplitude of the innermost power-law was fixed in such a way that the ``break'' in the $\xi(s)$ shape was at  $9 \Mpch$. The parameters characterizing the two power-laws are: $s_{0}=5.673 \Mpch$ and $\gamma = 1.30^{+0.06}_{-0.07}$, for $s<9 \Mpch$ and $s_{0}=7.5^{+0.4}_{-0.3} \Mpch$ and $\gamma = 3.29^{+0.21}_{-0.31}$, for $s>9 \Mpch$. This function is represented by the solid line, in  Fig. \ref{fig:xis}.

The shape of $\xi(s)$ suggests significant deviations from a simple power-law model. The interpretation of this result could include the effects of $z$-space distortions, affecting a power-law $\xi(r)$, whose form is derived from the $w_{p}(\sigma)$ results, as the small-scale random motions of the QSOs lead to a deficit of clustering amplitude measured at small scales. To confirm this hypothesis, we derived a $\xi(s)$ model from the best fitting power-law $\xi(r)$ to the $w_{p}(\sigma)$ results, by adding the distortions parameterized by $\beta(z=3)=0.25$ and $<w_{z}^{2}>^{1/2}=400 \kms$. These are likely values for those parameters (e.g., see \citet{fouc} and \citet{adel}). To include the distortions in $\xi(r)$ and derive a prediction for $\xi(s)$, a $\xi(\sigma,\pi)$ model was derived from the $\xi(r)$ input form, and then this was integrated in annuli to obtain $\xi(s)$. The $\xi(\sigma,\pi)$ model and this method are described in detail in \citet{me} (Model 1). The dotted black line in Fig. \ref{fig:xis} represents the obtained $\xi(s)$ result. This is still a good description of the data, within the errors, as it can be seen not only from the graph but also from the value of the reduced $\chi^{2}_{min}$, which is 1.25, for 16 degrees of freedom. Given that these are high-redshift galaxies, the $z$-space distortions in $\xi(s)$ caused by the small-scale random motions are not very significant, and hence the observed flattening of $\xi(s)$ can hardly be explained by $z$-space distortions only.\\

Results from several other galaxy and QSO surveys, such as the 2dF Galaxy and QSO Redshift surveys, have indicated the possible existence of a shoulder, at $\sim 8 - 12 \Mpch$, in the respective correlation functions (see \citet{hawk}, \citet{me}). These results are also backed by the shape of the correlation function as suggested from CDM model predictions, or Halo Occupation distribution (HOD) models (\citet{me}, \citet{tinker}). 

The double power-law model fitted to the data is a very good representation of the $\xi(s)$ results (the reduced $\chi^{2}_{min}$ of the fit is 0.66). The fact that the single power-law model for $\xi(r)$ also represents a good fit to the data, means that we can not prefer the double power-law model solely from the results of this work. However, when we take into account results from other surveys and theoretical models, the double power-law $\xi(s)$ model is more likely to be closer to the correct shape. Future LBG surveys will probe the projected correlation function at large separations. Even if these results, just from de-projection of $w_{p}(\sigma)$, do not lead to a conclusive measurements of the real-space correlation function, then their combination with large-scale statistically reliable measurements of $\xi(s)$ will make it possible to thoroughly study the $\xi(r)$ shape, strongly motivating future wide field spectroscopic surveys of LBGs. 

The double power-law $\xi(s)$ model can therefore be used as the input for the amplitude of the $z$-space correlation function, provided that the fit is only performed on scales where the non-linear distortions caused by the random peculiar motions have a negligible contribution. As it can be seen from the dotted black line, which represents the $\xi(s)$ obtained from a simple power-law $\xi(r)$ model, this is clearly not a problem at scales greater than $6 \Mpch$, where the function follows very closely a power-law form (assuming the value $<w_{z}^{2}>^{1/2}=400 \kms$).

\section{Constraints on $\beta$ and $\Omega_{m}^{0}$ from redshift-space distortions}

Once a model describing how the amplitude of the correlation function varies with radial separation, in average, then the higher-order distortions observed on $\xi(\sigma,\pi)$ can be modeled. 

The first step in modeling the $z$-space distortions is to build a $\xi(\sigma,\pi)$ model given an input $\xi(r)$. The latter can be obtained from a $\xi(s)$ form ($\gamma = 1.30^{+0.06}_{-0.07}$, for $s<9 \Mpch$; $\gamma = 3.29^{+0.21}_{-0.31}$, for $s>9 \Mpch$), if the $\xi(\sigma,\pi)$ fitting will be performed at scales where the main contribution for the distortions comes from the coherent infall, and hence the input $\xi(r)$ can be obtained by: 

\be
\xi(r) = \frac{\xi(s)}{1+\frac{2}{3}\beta(z)+\frac{1}{5}\beta(z)^{2}}
\label{equation:xisxir_kaiser}
\ee

The description of the $z$-space distortion model used here is present in section 7.3 of \citet{me}. Fitting this model to the measured $\xi(\sigma,\pi)$ will allow constraints on $\beta(z=3)$ to be drawn, provided that a given value for the velocity dispersion is assumed.

Constraints on the density parameter $\Omega_{m}^{0}$ are obtained from including geometric distortions in the $\xi(\sigma,\pi)$ model and comparing these with the distortions present in $\xi(\sigma,\pi)$ measured from the data. It is useful to make some definitions before continuing describing the cosmology fitting through geometric distortions. Following \citet{fiona02}, let the true, underlying cosmology of the Universe be the {\it true cosmology}, the cosmology used to build the model $\xi(\sigma,\pi)$ the {\it test cosmology}, and the cosmology assumed, both in the model and the data, to measure the correlation function, the {\it assumed cosmology}.\\

In the linear regime, the correlation function in the assumed cosmology will be the same as the correlation function in the test cosmology, given that the separations are scaled adequately. For $\xi(\sigma,\pi)$:

\be
\xi_{t}(\sigma_{t},\pi_{t}) = \xi_{a}(\sigma_{a},\pi_{a}),
\label{equation:xisp_cosmol}
\ee
where the subscripts $t$ and $a$ refer to the test and assumed cosmology, respectively.\\

The relation between the separations $\sigma$ and $\pi$ in the two cosmologies is the following \citep{bph}, \citep{fiona00}:

\be
\sigma_{t} = f_{\perp} \sigma_{a} = \frac{B_{t}}{B_{a}}\ \sigma_{a}
\label{equation:sig_cosmol}
\ee
\be
\pi_{t} = f_{\parallel} \pi_{a} = \frac{A_{t}}{A_{a}}\ \pi_{a}
\label{equation:pi_cosmol}
\ee
where $A$ and $B$ are defined as follows:

\be
A = \frac{c}{H_{0}}\frac{1}{\sqrt{\Omega_{\Lambda}^{0}+\Omega_{m}^{0}(1+z)^{3}}}
\label{equation:A_cosmol}
\ee

\be
B = \frac{c}{H_{0}}\int_{0}^{z}\frac{dz'}{\sqrt{\Omega_{\Lambda}^{0}+\Omega_{m}^{0}(1+z')^{3}}}
\label{equation:A_cosmol}
\ee
if the both true and assumed cosmologies correspond to spatially flat Universes.

If the same cosmology is assumed in the data and the model, then the observed shapes of the measured and modeled $\xi(\sigma,\pi)$ will be the same when the test cosmology is the same as the true, underlying cosmology of the Universe. Therefore, constraints on $\beta(z=3)$ and $\Omega_{m}^{0}$ are obtained from simply doing a $\chi^{2}$ fit between the data and the a series of different $\xi(\sigma,\pi)$ models, derived with different test cosmologies and values of $\beta(z=3)$. 

The steps taken for this fitting procedure are as follows:

\begin{itemize}

\item Assume a cosmology and measure $\xi(s)$, $w_{p}(\sigma)$, $\xi(\sigma,\pi)$.

\item Take a model for the $z$-space correlation function, e.g. a double power-law. This model should be a good description of the observed data at those scales not significantly affected by velocity dispersion and included in the $\xi(\sigma,\pi)$ fit.

\item Choose a pair of test values of $\Omega_{m}^{0}$ and $\beta(z)$.

\item The model for $\xi(s)$ is a good description for the data in the assumed cosmology. What is actually needed at this stage is a $\xi(\sigma,\pi)$ model in some test cosmology, hence the correct input for this model is $\xi(s)$ in that same test cosmology. Since, in the linear regime, $\xi_{t}(r_{t}) = \xi_{a}(r_{a})$, one has only to compute the real-space separation in the assumed cosmology to get $\xi_{t}(r_{t})$. The relation between the real- and $z$-space correlation functions in any cosmology is given by Eq. \ref{equation:xisxir_kaiser}; $r_{t}$ is given by $r_{t} = \sqrt{\sigma_{t}^2+(\pi_{t}-w_{z}/H_{t})^{2}}$ and the relation between $r_{a}$ and $r_{t}$ is: $r_{a} = r_{t}/(f_{\perp}^{2}f_{\parallel})^{1/3}$.

\item Using that model for $\xi_{t}(r_{t})$, compute $\xi_{t}(\sigma_{t},\pi_{t})$. Then, include the geometric distortions by scaling $\xi_{t}(\sigma_{t},\pi_{t})$ back to the assumed cosmology, in a similar way as described in the previous step. To get $\xi_{a}(\sigma_{a},\pi_{a})$, one needs to scale the separations $\sigma_{t}$ and $\pi_{t}$ to $\sigma_{a}$ and $\pi_{a}$, using equations \ref{equation:sig_cosmol} and \ref{equation:pi_cosmol}.

\item Adding the effects of large-scale infall not only introduces
  distortions in $\xi(\sigma,\pi)$ but also shifts the amplitude of the
  correlation function, by an amount that depends on the value of
  $\beta(z)$ taken. Since the amplitude of the spherical-averaged
  $z$-space correlation function is introduced as an input to the model, it always remains the same, whatever $\beta$ and $\Omega_{m}^{0}$ are used as {\it test}
  values. This guarantees that the fit is being made to the {\it distortions} in $\xi(\sigma,\pi)$, since the averaged {\it amplitude} remains the same for whatever combination of $\beta$ and $\Omega_{m}^{0}$.

\item For the best fitting value of this amplitude factor, determine
  the $\chi^{2}$ value for the fit of this model to the data.

\item Repeat this procedure for a different combinations of $\Omega_{m}^{0}$ and $\beta(z)$.

\end{itemize}

The number of degrees of freedom in the $\chi^{2}$ fit is the total
number of bins where $\xi(\sigma,\pi)$ from the model is fitted to the
data minus the number of free parameters. If the fit is to
$\Omega_{m}^{0}$ and $\beta(z)$, the number of free parameters will be two. Although it can be argued that the $\xi(\sigma,\pi)$ bins may not be independent, the accuracy of these errors is supported by $N$-body simulations \citep{fiona00}. The velocity dispersion was fixed to $400 \kms$ \citep{adel}.

\begin{figure}
\begin{center}
\centerline{\epsfxsize = 9.0cm
\epsfbox{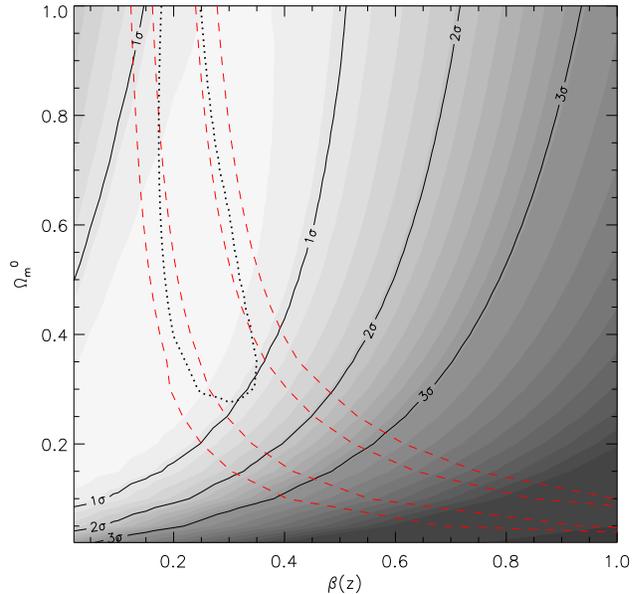}}
\caption{The confidence levels on the $[\Omega_{m}^{0}, \beta(z)]$ plane from fitting the $z$-space distortions in $\xi(\sigma,\pi)$ (grey scale and solid lines). The dashed lines show the $1 \sigma$ and $2 \sigma$ confidence levels obtained from linear growth theory, considering the value of $\beta$ for local galaxies surveys. The joint $1 \sigma$ confidence level is given by the dotted line.}
\label{fig:xisp_fit}
\end{center}
\end{figure}

The result of doing this fit in the present data is shown in Fig. \ref{fig:xisp_fit}.

\begin{figure}
\begin{center}
\centerline{\epsfxsize = 5.0cm
\epsfbox{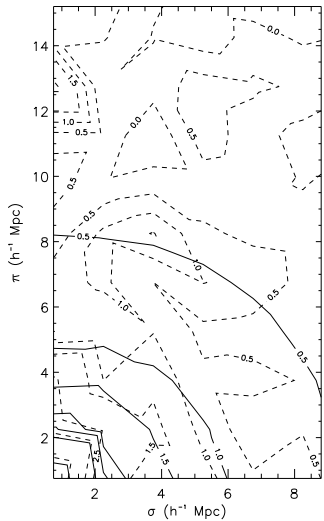}}
\caption{The best fitting model for $\xi(\sigma,\pi)$, obtained using the best fitting values of  $\beta(z=3) = 0.25^{+0.05}_{-0.06}$ and $\Omega_{m}^{0} = 0.55^{+0.45}_{-0.16}$ (solid line) and $\xi(\sigma,\pi)$ measured from the data (dashed line).}
\label{fig:xisp_data_model}
\end{center}
\end{figure}

The shaded regions and the solid line refer to the confidence levels obtained from fitting the $z$-space distortions. These constraints alone correspond to best fitting values of  $\beta(z=3) = 0.15^{+0.20}_{-0.15}$ and $\Omega_{m}^{0} = 0.35^{+0.65}_{-0.22}$. The dashed lines are the $1 \sigma$ and $2 \sigma$ confidence levels obtained from clustering evolution, from applying linear growth theory and predicting the values of $\beta(z=3)$ for different cosmologies, using as an input the value of $\beta$ at $z\sim 0.1$, obtained from the 2dFGRS survey \citep{hawk}. This method is described in detail in \citet{me}. The dotted line represents the $1 \sigma$ two parameter joint confidence level. The best fitting values are $\beta(z=3) = 0.25^{+0.05}_{-0.06}$ and $\Omega_{m}^{0} = 0.55^{+0.45}_{-0.16}$. Fig. \ref{fig:xisp_data_model} shows the $\xi(\sigma,\pi)$ data (dashed line) and the best fitting $\xi(\sigma,\pi)$ model (solid line), obtained from the joint constraints on $\beta(z=3)$ and $\Omega_{m}^{0}$. A visual comparison between the solid and the dashed contours also shows that the model is a good description of the $\xi(\sigma,\pi)$ data.

\section{Discussion and Conclusions}

Here we have used the $z$-space distortions on the clustering pattern of a sample of LBGs to derive constraints on $\beta(z=3)$ and $\Omega_{m}^{0}$. The method used requires the spherical average amplitude of the correlation function to be accurately known and given as an input to the model. Due to the small size of the fields used in the survey, the clustering measured across the line-of-sight is not probed at scales larger than $10 \Mpch$. For this reason, and given that at $z=3$ only the smallest scales are significantly affected by non-linear distortions caused by the random motions of the galaxies, the input amplitude of the spherically averaged correlation function should be obtained from $\xi(s)$. 

The $\xi(s)$ shape suggests that the real-space correlation function of the LBGs might deviate significantly from a simple power-law model, and these deviations are not evident in the projection $w_{p}(\sigma)$. Future LBG surveys where larger scales across the line-of-sight are probed, will allow a coherent picture of the clustering to be drawn from the $w_{p}(\sigma)$ and $\xi(s)$ measurements, similarly to what has been done for the 2dF QSO Redshift Survey \citep{me}. 

Once a suitable model for describing the amplitude of the correlation function as a function of scale is obtained (a two power-law model was found to be a good description in the present case), then the $z$-space and geometric distortions can be modeled and constraints on $\beta(z=3)$ and $\Omega_{m}^{0}$ drawn. The combination of these constraints with orthogonal confidence levels from linear growth of density perturbations, indicates that $\beta(z=3) = 0.25^{+0.05}_{-0.06}$ and $\Omega_{m}^{0} = 0.55^{+0.45}_{-0.16}$. These values are consistent with previous measurements of the bias. \citet{fouc} found, for a $\Lambda$CDM cosmology, $b=3.5\pm 0.3$, using a sample of LBGs from the Canada-France Deep Field Survey. Considering the WMAP results \citep{wmap}, if we take $\Omega_{m}^{0}=0.3$, which is within our computed error bars, their obtained value of $b$ corresponds to $\beta(z=3)=0.27$, which is well within our derived error bars and hence consistent with our results. However, results from our $\xi(\sigma,\pi)$ fits strongly relies on the amplitude and shape of the input correlation function. If our input $\xi(s)$ shape is uncertain, then this may introduce errors in the measured constraints of $\Omega_{m}^{0}$ and $\beta(z=3)$. We found that the derived constraints are very robust to $1\ \sigma$ changes in the parameters describing the double power-law $\xi(s)$ model. However, changes to the assumed $\xi(s)$ (or $\xi(r)$) shape, will lead to more significant changes in the derived $\Omega_{m}^{0}$ and $\beta(z=3)$ constraints. Unfortunately, with the present data it is difficult to measure the real-space correlation function. This handicap is basically due to two factors: the size of fields used; and the number of LBGs in the survey. Even though the latter is, in part, balanced by the high space density of these galaxies, the former prevents any real-space clustering measurement at scales $\gsim 10 \Mpch$, except via the $\xi(s)$ form. This hampers any attempt to build such a coherent picture of LBG clustering. We believe we have done the best that is possible with the current data. It is important to stress the advantage of the LBGs high spatial density for $z$-space distortion analyses. The constraints in $\Omega_{m}^{0}$ and $\beta(z)$ from the $\xi(\sigma,\pi)$ fitting alone are comparable to those achieved with the 2QZ sample \citep{me}, which includes $\sim 20000$ QSOs. However, the much smaller number of objects, in the present work, is counterbalanced by their high-spatial density, which dramatically increases the clustering signal. Our results not only reflect what can be achieved in cosmological and dynamical constraints from $z$-space distortions, but they also foreshadow future work, based on larger LBG surveys.

%Unfortunately, with the present data it is not possible to create a statistically reliable and coherent picture, where we present measurements of $w_{p}(\sigma)$, $\xi(r)$, $\xi(s)$, $\Omega_{m}^{0}$ and $\beta(z=3)$. 

%The constraints derived here for $\Omega_{m}^{0}$ are not competitive with other current methods. However, based on the {\it Hubble Volume} simulations \citep{jenkins} we found that with a survey of $\sim 2000$ LBGs, and probing scales in the sky direction achievable with the VIMOS instrument at the VLT, the present method allows the determination of $\Omega_{m}^{0}$ with an uncertainty already of $\sim 0.1$. This therefore suggests that this method, applied to large future surveys of LBGs, will allow new, tight constraints on cosmology and the bias to be derived.

\section*{Acknowledgements}

JA acknowledges financial support from FCT/Portugal
through project POCTI/FNU/43753/2001 and from the European Community's
Human Potential Program under contract HPRN-CT-2002-00316,
SISCO. PJO acknowledges the support of a PPARC Fellowship. We thank Catarina Lobo for useful comments.

{}

\label{lastpage}

\end{document}